# Optical detection of Aflatoxins B in grained almonds using fluorescence spectroscopy and machine learning algorithms

F. R. Bertani[a], L. Businaro[a], L. Gambacorta[b], A. Mencattini[c], D. Brenda[c], D. Di Giuseppe[c], A. De Ninno[a], M. Solfrizzo[b], E. Martinelli[c], A. Gerardino[a]

[a]Consiglio Nazionale delle Ricerche, Istituto di Fotonica e Nanotecnologie, via Cineto Romano 42, 00156, Roma, Italia

[b] Consiglio Nazionale delle Ricerche, Istituto di Scienze delle Produzioni Alimentari, via Amendola, 122/O - 70126 Bari, Italia

[c]Department of Electronic Engineering, University of Rome Tor Vergata, via del Politecnico 1, 00133, Roma, Italia

e-mail: francesca.bertani@ifn.cnr.it

**Abstract**

Aflatoxins are fungal metabolites extensively produced by many different fungal species that may contaminate a wide range of agricultural food products. They have been studied extensively because of being associated with various chronic and acute diseases especially immunosuppression and cancer and their presence in food is strictly monitored and regulated worldwide.

Aflatoxin detection and measurement relies mainly on chemical methods usually based on chromatography approaches, and recently developed immunochemical based assays that have advantages but also limitations, since these are expensive and destructive techniques. Nondestructive, optical approaches are recently being developed to assess presence of contamination in a cost and time effective way, maintaining acceptable accuracy and reproducibility. In this paper are presented the results obtained with a simple portable device for nondestructive detection of aflatoxins in almonds. The presented approach is based on the analysis of fluorescence spectra of slurried almonds under 375 nm wavelength excitation. Experiments were conducted with almonds contaminated in the range of 2.7-320.2 ng/g total aflatoxins B (AFB$_1$ + AFB$_2$) as determined by HPLC/FLD. After applying pre-processing steps, spectral analysis was carried out by a binary classification model based on SVM algorithm. A majority vote procedure was then performed on the classification results. In this way we could achieve, as best result, a classification accuracy of 94% (and false negative rate 5%) with a threshold set at 6.4 ng/g. These results illustrate the feasibility of such an approach in the great challenge of aflatoxin detection for food and feed safety.

1. Introduction

Aflatoxins are a group of chemically related mycotoxins formed by microscopic fungi (*Aspergillus flavus*, *A. parasiticus*, and *A. nomius*) occurring in corn, cottonseed, peanuts and other nuts, grains and spices (Eaton and Groopman, 1993; Fao, 2016) Fungal infection and aflatoxin production can occur at any stage of plant growth, harvesting, drying, processing, and storage. Both the infection process and aflatoxin accumulation are strongly affected by environmental conditions such as insect damage, temperature, and humidity. Moreover, aflatoxins can contaminate the food both before and after harvest and cannot be destroyed by any form of food processing (Rustom, 1997). Aflatoxins occur in more than ten varieties (including B1, B2, G1, and M1 and M2 which are secreted in the milk of lactating animals that have consumed feed contaminated with aflatoxin) , of which  aflatoxin B1 is the most prevalent and toxic one (Rasch et al., 2010). Exposure by ingestion or inhalation of aflatoxins may lead to the development of serious medical conditions that vary considerably depending on the animal species, dose, diet, age, and gender(Kumar et al., 2017). Acute effects are primarily observed in structural and functional damage of the liver, including liver cell necrosis, haemorrhage, lesions, fibrosis, and cirrhosis. Additionally, hepatic encephalopathy, immunosuppression, lower respiratory infections, anorexia, malaise, and fever have been observed. Is furthermore considered the most potent naturally occurring carcinogen and is designated group 1 human carcinogen (Eaton and Groopman, 1993; IARC, 1993; Streit et al., 2012).

The maximum permitted levels of aflatoxins in food and feed products are regulated in over 100 nations (Unnevehr and Grace, 2013). The European Commission for example states the maximum allowed total aflatoxin levels in different food materials  (Commission Regulation (EC), 2006a), while the USA food safety regulations included a limit of 20 ng/g of total aflatoxins in all food products (FDA, 2000). There are currently several different accepted measurement techniques available for determining aflatoxin levels

(Fao, 2016; Ricci et al., 2007). In general, conventional measurement techniques require three steps: extraction with mixture of water and organic solvents; purification of crude extract to concentrate the extract and remove interferents as much as possible; finally separation, detection and quantification. Conventional purification techniques are based on solid phase extraction (SPE), a mainstay of analytical chemistry. Examples include, reversed phase- and silica-based SPE columns and immunoaffinity based columns. For separation, detection and quantification high pressure liquid chromatography (HPLC) coupled with fluorescence or mass spectroscopy (MS) detector is mainly used for food and feed analysis whereas thin layer chromatography (TLC) is mainly used in developing countries (Fao, 2016). HPLC based methods are well-proven and widely accepted, however, they are laborious, time consuming, require well trained personnel and are costly because they need a significant investment in consumables, equipment and maintenance (CEN-EN, 2007). Some more modern and rapid methods (Yao et al., 2015) are semiquantitative such as ELISA, lateral flow tests, direct fluorescence, fluorescence polarization immunoassay and biosensors (Dinçkaya et al., 2011; Eivazzadeh-Keihan et al., 2017). Indirect methods are based on mid- or near-infrared spectroscopy (Kimuli et al., 2018; Pearson et al., 2001; Wu et al., 2018) and do not directly measure aflatoxins but measure an indicator that is associated with the production of aflatoxins. Emerging technologies include hyperspectral imaging (Kalkan et al., 2011; Wang et al., 2015), electronic nose, aptamer-based biosensors (Castillo et al., 2015), molecular imprinted polymers, nanoparticle-based and cell-based approaches (Larou et al., 2013).

Besides being costly, destructive and time consuming, the sample-based analysis often gives a limited view on the degree of contamination if the laboratory sample is not obtained with a correct sampling procedure (Commission Regulation (EC), 2006b; Commission Regulation (EU), 2014). This is due mainly to the inhomogeneity of the toxin distribution in both the food products and the crops. The true aflatoxin level cannot be determined with 100% accuracy, causing difficulties in the segregation of aflatoxin-contaminated food and feed without destroying a large amount of products and incurring significant economic losses (Shanakhat et al., 2018). The current published non-invasive and non-destructive optical (Tao et al., 2018) and fluorescence methods allow the identification of toxins in liquids, like beer or wine, and in grains and nuts (Rasch et al., 2010; Smeesters et al., 2015). In addition, the fluorescent toxins can be easily identified if no or very low background fluorescent elements are present. Cereals and nuts usually contain various fluorescent proteins which make the aflatoxin detection difficult (Schade and King, 1984). Nevertheless, because aflatoxins are fluorescent substances, fluorescence spectroscopy seems the most promising non-destructive optical detection technique. To detect the aflatoxins, recent studies focused on analyzing both the one- and two-photon induced fluorescence spectra of healthy and contaminated maize grains (Smeesters et al., 2015) employing huge laboratory instruments (lasers, sensors, etc).

In our experiments we found that almond matrices show a marked intrinsic fluorescence emission in the observed spectral range. Despite this intrinsic fluorescence signal from matrices, AFB detection in the range of the concentration analyzed (ng/g) is possible by fluorescence spectra analysis and machine learning algorithms implementation. In this paper preliminary results of fluorescence spectroscopy for the detection of aflatoxin contamination on slurried almonds are presented. In future, the availability of a multispectral, cheap, rapid and non-destructive technique will represent a real benefit in the field of food safety, especially in developing countries or in rural environments. Detection of mycotoxins in food products is one of the objectives of the PhasmaFOOD project (www.phasmafood.eu), funded by the EU program H2020, that has the objective to develop a multi-sensing platform by the integration of VIS and NIR micro-spectrometers that will serve as actual chemical sensor.

## 2. Material and methods

### 2.1 Production of aflatoxin contaminated almonds

Shelled almonds (cs. Genco) were inoculated with a strain of *Aspergillus flavus* (ITEM 7828) that produces aflatoxins $B_1$ and $B_2$, according to the experimental conditions reported elsewhere (Zivoli et al., 2014). After incubation with *A. flavus*, almonds (1 kg) were dried at 40 °C for 48 hr, ground, homogenized and (3.12 g) mixed with 496.97 g of blank (uncontaminated) ground almonds. The 500 g of contaminated almonds were added with 1000 ml water and slurried for 5 min with a T25 ULTRA-TURRAX (IKA®-Werke GmbH & CO. KG, Staufen im Breisgau, Germany) to obtain ground almonds homogeneously contaminated with aflatoxins. The slurried almond sample was freeze-dried, ground and analysed in triplicate by HPLC-FLD (see below) resulting in $AFB_1$ and $AFB_2$ levels of 290.6 ± 9.0 ng/g and 29.2 ± 1.6 ng/g, respectively. Different aliquots of this contaminated sample were mixed with blank ground almonds to prepare 9 x 80 g of ground almonds containing lower levels of $AFB_1$ and $AFB_2$. The eight contaminated almond samples were each slurried for 5 min with 160 ml water, freeze-dried, ground an and analysed by HPLC-FLD to accurately measure the levels of $AFB_1$ and $AFB_2$. Samples of blank ground almond and contaminated ground almonds were then analysed in quadruplicate by fluorescence spectroscopy.

### 2.2 Determination of $AFB_1$ and $AFB_2$ in contaminated almonds

A validated HPLC-FLD method (Zivoli et al., 2014) was used to accurately measure aflatoxin levels in artificially contaminated samples of ground almonds. In particular, 5 g of dry ground almonds were extracted with 50 ml of a mixture of acetone/water (85:15 v/v) by sonication for 30 min. After filtration on fluted no. 4 filter paper (Whatman), 5 ml of filtered extract was diluted with 75 ml of ultrapure water and filtered with a GF/A glass microfiber filter (Whatman). Forty milliliters of filtered diluted extract (equivalent to 0.25 g of matrix) was passed through the immunoaffinity column AflaTest (Vicam, Milford, MA). The column was then washed twice with 10 ml of ultrapure water, and aflatoxins were eluted with 1.5 ml (3 × 0.5 ml) of methanol. The three methanolic eluates were collected, diluted with ultrapure water up to 5 ml in a volumetric flask, and analyzed by HPLC-FLD. The samples extracts containing high aflatoxin concentration that overloaded the capacity of the immunoaffinity were reanalyzed. In particular, the samples extract was diluted ten times with water and 5 ml were purified as described above.

The HPLC-FLD analyses were performed with an Agilent 1260 Infinity (AgilentTechnologies, Inc., Wilmington, DE, USA) consisting of a binary pump (G1312B), an autosampler (G1367E) with a 100 µl loop, a fluorescence detector (G1321B) fixed at 365 nm $\lambda_{ex}$ and 435 nm $\lambda_{em}$, a column oven (G1316C) set at 30 °C and a software for Microsoft Windows 7 (OpenLAB, CSB, ChemStation Edition). The column used was a 150 mm × 4.6 mm i.d., 3 µm, Luna PFP (2) (pentafluorophenyl-propyl) with a 3 mm i.d., 0.45 µm pore size guard filter. The chromatographic separation was performed in the isocratic condition using a mixture of MeCN/$H_2O$ (30:70 v/v) at flow rate of 0.8 ml/min. A photochemical post-column derivatization UVE system (LCTech, Dorfen, Germany) was used to enhance the fluorescence of $AFB_1$. The results of $AFB_1$ and $AFB_2$ levels measured in the 9 samples of ground almonds are reported in Table 1.

| Sample | AFB$_1$ (ng/g) | AFB$_2$ (ng/g) | AFB$_1$+AFB$_2$ (ng/g) |
|---|---|---|---|
| Almond 1 (control, uncontaminated) | ND | ND | - |
| Almond 2 | 291.0 ± 8.7 | 29.2 ± 1.5 | 320.2 ± 8.9 |
| Almond 3 | 39.3 ± 1.2 | 5.9 ± 0.3 | 45.2 ± 1.2 |
| Almond 4 | 16.4 ± 0.5 | 1.9 ± 0.1 | 18.3 ± 0.5 |
| Almond 5 | 12.5 ± 0.4 | 1.2 ± 0.06 | 13.7 ± 0.4 |
| Almond 6 | 9.9 ± 0.3 | 0.8 ± 0.04 | 10.7 ± 0.3 |
| Almond 7 | 8.3 ± 0.2 | 0.6 ± 0.03 | 8.9 ± 0.3 |
| Almond 8 | 6.0 ± 0.2 | 0.4 ± 0.02 | 6.4 ± 0.2 |
| Almond 9 | 4.2 ± 0.1 | 0.2 ± 0.01 | 4.4 ± 0.1 |
| Almond 10 | 2.6 ± 0.1 | 0.1 ± 0.01 | 2.7 ± 0.1 |

Table 1. Levels of AFB$_1$ and AFB$_2$ in samples of ground almonds used in this study. ND: not detected (LOD 0.2 ng/g)

### 2.2 Rapid fluorescence measurement

The experimental dataset has been collected through two different measurement sessions, the first at location 1 (ISPA-CNR laboratories of Bari, Italy) in April 2018, the second at location 2 (Rikilt laboratories in Wageningen, Netherlands) in June 2018.

To perform the fluorescence spectroscopy measurements, a commercial Hamamatsu micro-spectrometer model C12880MA has been identified for its characteristics of sensitivity and wide spectral range (340-850 nm). The micro-spectrometer has been mounted in a 90 degree configuration with the lighting system, in a optical set-up that ensures a fixed distance of 5 cm from sample surface to spectrometer, when the instrument is kept vertically on the edge of the dish containing the sample (Fig. 1).

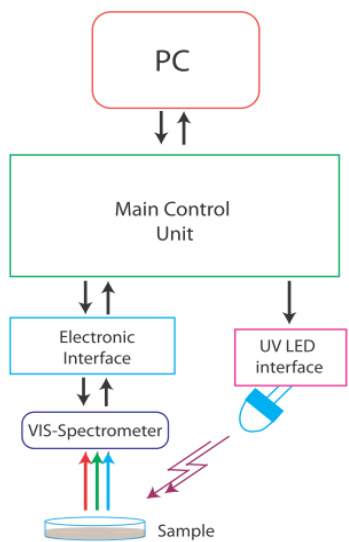

Figure 1 Scheme of the fluorescence spectroscopy set-up.

The illumination system consists of two UV LEDs (Nichia radial 405648, 375 nm wavelength emission, radiant flux typically 7500 µW). Hamamatsu electronic evaluation board and acquisition software have been used to drive the micro-spectrometer and collect the data. In order to avoid saturation effects from excitation light beam reflectance, a longpass plastic sheet filter with 400 nm cut-on wavelength has been placed in front of spectrometer entrance slit (39-426 Edmund Optics).

To collect the spectra, the integration time has been set at 200 ms and a dark signal acquired with UV excitation light off. A darkening shield has been used to avoid environmental light contributions.

In order to take into account variability in the measurement procedures, measurements have been acquired in two different laboratories and in different days for each laboratory session: for each of the ten contamination levels (1 non-contaminated batch and 9 other contaminated batches containing levels of 2.7, 4.4, 6.4, 8.9, 10.7, 13.7, 18.3, 45.2, 320.2 ng/g of total aflatoxins), quadruplicate samples have been prepared putting ground almond in 60 mm-diameter Petri dishes. Moreover, in the two measurement sessions (locations 1 and 2), it has not been used the same instrument but two nominally identical instruments mounting the same elements (C12880MA spectrometer and Nichia LEDs) in the same configuration. UV-excited fluorescence spectra have been recorded in five different spatial positions over each sample, and for each position five replicates have been acquired and averaged. Therefore, almond dataset consists of 1000 measures per day, 2000/3000 measures per session and 5000 total measures.

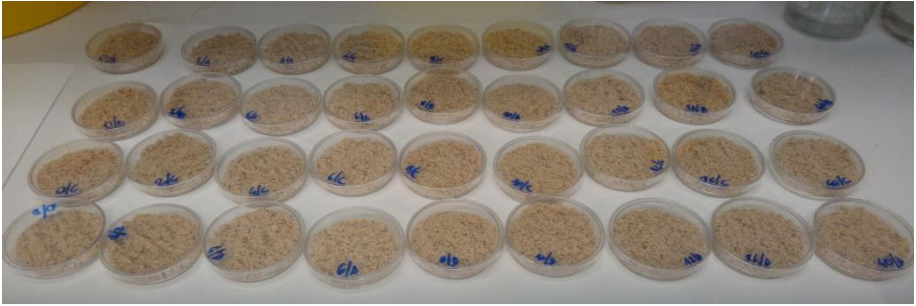

Figure 2: Set of samples. In the figure the level from 0 to 45.6 ng/g are shown in quadruplicate.

## 2.3 Data analysis

The data analysis process has been divided into two consecutive phases, a preliminary analysis made of a set of pre-processing steps and the classification procedure itself. A schematic picture of the method is shown in Fig.3.

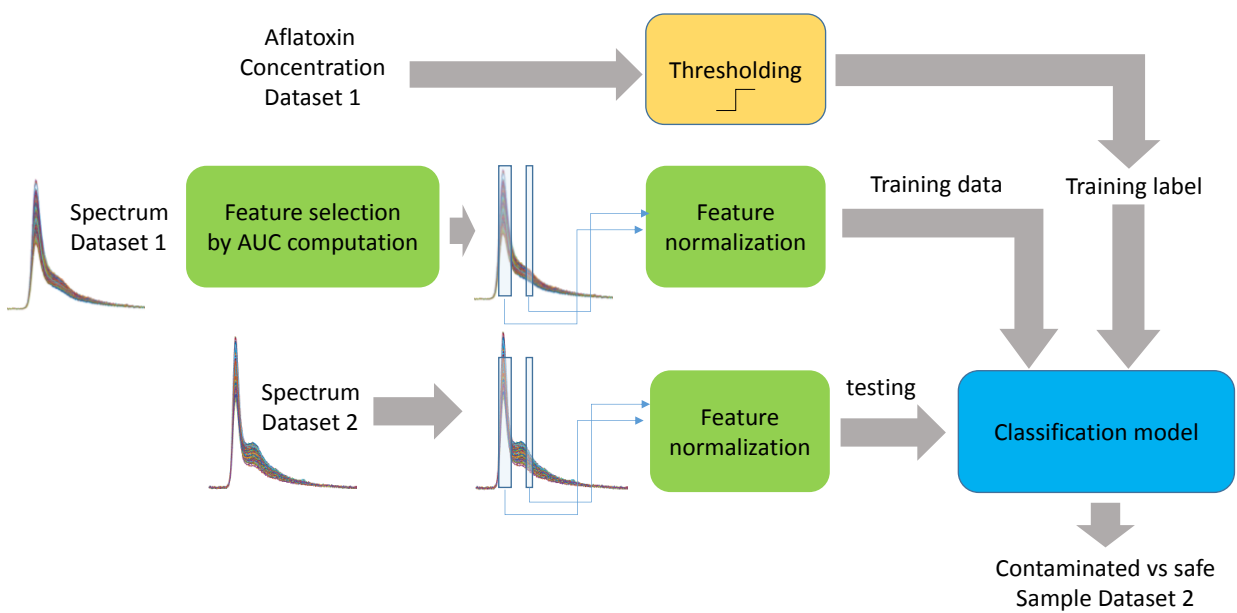

Figure 3. A schematic flowchart of the whole method for the data analysis. Spectrum dataset 1 is used for feature selection. Selected features are all normalized in training and testing by autoscaling. The reference value of aflatoxin is then thresholded in order to assign a label to the training features of safe (0) vs contaminated (1). The selected level for thresholding is the same for training and testing dataset. A classification model is then constructed on the training set and applied to the testing dataset for sample recognition. The classification step includes a majority voting procedure aimed at performing the majority voting procedure among labels assigned to the testing dataset.

The preliminary analysis has been structured into a set of pre-processing steps, whose main target was to build the two datasets (one for each measurement session) and to identify the optimal threshold value for class definition purpose. In Fig. 4, the spectrum of 18.3 ng/g samples is shown and a peak at about 410 nm is present, due to the not totally filtered excitation reflectance signal. This peak is present in all the collected spectra and represents a common characteristic of all the data.

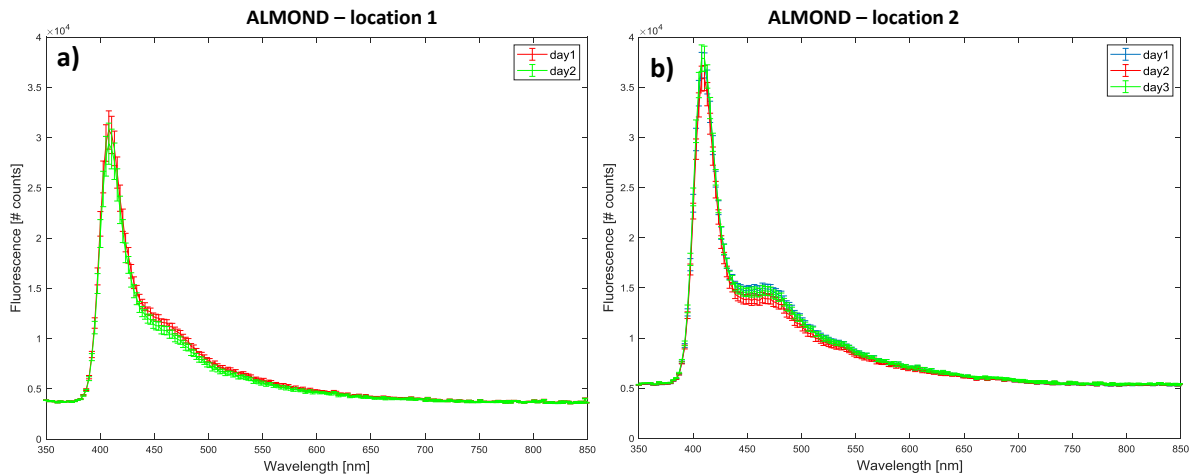

Figure 4. Data collected in location 1 (a) and location 2 (b) (different instruments) in different days on the same sample (18.3 ng/g). Mean and standard deviation values over the same samples are shown. Similar behavior is shown by the fluorescence spectra collected on the other samples.

Collected data resulted to have a moderate reproducibility between-day (see Fig. 4) and a relevant dispersion of data. This effect is mainly due to the differences in the experimental setups. The proper use of the information contained in the spectra during the data analysis phase and the creation of appropriate datasets required the application of a feature selection and normalization strategy for the fluorescence spectra. In the proposed strategy, for selecting features, we computed the Area Under the Receiving Operating Characteristic (ROC) Curve (AUC) score (Bradley, 1997) for each spectrum dataset with respect to the corresponding reference contamination level. The procedure is applied on the training set. The AUC is computed for each spectrum value and features exhibiting AUC values higher than 0.6 are then selected and used also in the testing set. Normalization has been performed by standard autoscaling (Mclachlan, 1992).

For the classification task, we applied a Support Vector Machine (SVM) algorithm with Radial Basis Function kernel to the binary classification problem of safe vs contaminated samples (C. Bishop, 2009).

For the sake of investigation, we considered different threshold values between 2.7 and 13.7ng/g. The range has been chosen according to the EU regulation limits (8-10 ng/g in the case of almond(CEN-EN, 2007)). For each threshold value $TH$, we define the two classes $C_1$ and $C_2$ as

$$C_1 \rightarrow L_{cont.} \leq TH$$

$$C_2 \rightarrow L_{cont.} > TH$$

Where $L_{cont.}$ represents the actual contamination level of the considered sample, while $TH$ represents the threshold value. Independent dataset testing has been implemented by using a location for training and the other location for testing and viceversa.

Standard classification approach has been also supported by a majority voting procedure applied to the labels (classification results) assigned to the spectra acquired on samples with the same contamination level and in different spatial positions (see par. 2.2). In such a way, the approach exhibits increased robustness to heterogeneity of the sample contamination that may occur.

The parameters used to evaluate the goodness and the reliability of each result are represented by prediction accuracy. We define as instance the result of a measurement (fluorescence spectrum). The notation used to indicate the four types of instances within the confusion matrix is the following:

- TP (True Positive) → Highly contaminated instances that have been classified properly
- TN (True Negative) → NOT contaminated instances that have been classified properly
- FN (False Negative) → Highly contaminated instances that have been classified as NOT contaminated
- FN (False Positive) → NOT contaminated instances that have been classified as highly contaminated

According to this notation, the classification accuracy (ACC) can be calculated as follows:

$$ACC = \frac{TP + TN}{TP + TN + FN + FP}$$

Table 2 shows the accuracy values (ACC) and the False Negative Rate (FNR) (i.e., the number of false negatives achieved divided by the total number of positive instances, TP+FN) for the two locations: "Location 1" means that the data collected in location 1 have been used as test and the data collected in location2 has been used as training data, Location 2 is the reverse case. The first column shows the fixed contamination threshold; the next two columns display ACC and FNR obtained for training data set; the last four columns list the ACC and FNR for testing data set before and after the implementation of the majority voting procedure on the instances coming from samples with the same contamination level. The FNR value is crucial in food contamination analysis where the risk associated to a false negative instance (contaminated sample misclassified as safe) is higher than the risk of a false positive instance (safe sample misclassified as contaminated). As expected the optimal threshold value represents a trade-off condition. When the threshold value is low many more samples are assigned to the contaminated class. This situation may affect the performance of the training step and the model construction. As a consequence, the model is capable to better recognize contaminated rather than safe samples and the number of FNRs reduces. When the threshold value is high, a few samples are assigned to the contaminated category. Model capability to recognize contaminated samples reduces and also FNRs increase. Overall, the accuracy values obtained take into account both FNRs and FPRs and, as a consequence, represents a balancing of such kind of errors.

|  | Threshold (ng/g aflatoxins) | ACC training | FNR training | ACC testing | FNR testing | ACC Testing Majority voting | FNR Testing Majority voting |
|---|---|---|---|---|---|---|---|
| **Location 1 (test) Location 2 (training)** | 2 | 0.9890 | 0.0067 | 0.8605 | 0.0269 | 0.8875 | 0.0156 |
|  | 4 | 0.9917 | 0.0033 | 0.8275 | 0.0121 | 0.8500 | 0 |
|  | 6 | 0.9867 | 0.0078 | 0.7765 | 0.0425 | 0.8000 | 0 |
|  | 8 | 0.9480 | 0.0173 | 0.7200 | 0.1690 | 0.7875 | 0.0500 |
|  | 10 | 0.9263 | 0.0675 | 0.6755 | 0.3350 | 0.7500 | 0.2188 |
| **Location 2 (test) Location 1 (training)** | 2 | 0.9805 | 0.0025 | 0.8750 | 0.0742 | 0.8750 | 0.0729 |
|  | 4 | 0.9630 | 0.0050 | 0.8750 | 0.0638 | 0.8917 | 0.0476 |
|  | 6 | 0.9350 | 0.0108 | 0.8923 | 0.0994 | 0.9417 | 0.0556 |
|  | 8 | 0.9185 | 0.0360 | 0.8210 | 0.2813 | 0.8583 | 0.2167 |
|  | 10 | 0.9160 | 0.1175 | 0.7323 | 0.5833 | 0.7333 | 0.6458 |

Table 2. Results of classification for the two sessions in terms of ACC and FNR in training, in testing, and after the implementation of majority voting.

## Discussion

The results reported in Tab.2 gives evidence to diverse considerations. First of all, location1 and location2 experiments present, as above discussed, a moderate reproducibility in spectra, here represented by the different performance results. In particular, model trained on location 1 better recognizes contamination effects in dataset collected in location 2 (accuracy up to 0.94) with respect to the model trained on data acquired on location 2 (accuracy up to 0.88).

Secondly, the heterogeneity of the spectra acquired at different positions in samples with the same nominal contamination is partially encompassed by the majority voting strategy that produces an increase in the accuracy value and an acceptable decrease in the FNRs except for the threshold value of 10.7 ng/g. Beyond the previous considerations, an important result is the demonstration of the feasibility to select a unique threshold value for the classification of contaminated vs safe samples (binary classification model). The best results in terms of accuracy (94%) and false negative rate (5%) have been achieved using the location 2 dataset as test and location 1 data set as training and a threshold equal to 6.4 ng/g. This result must be framed in the specific experimental conditions in which the data have been collected: the reference samples have been prepared accordingly to standardised protocols and are quite homogeneous in granularity. The spectroscopy measurements have been performed in reproducible conditions and the number of spectra acquired per sample are in the range of thousands, also if different locations have been included. The same kind of analysis must be performed on a more populated data set and in "real setting" conditions to prove its effective feasibility.

As an explicative example, fixing the threshold value to 4.4 ng/g (indicating as contaminated samples the ones having a concentration of aflatoxin strictly larger than 4.4ng/g), the method results in an accuracy value of 0.85 and 0.89 for the two locations respectively. Correspondingly, we obtained an FNR of 0 and 0.0476 respectively for locations 1 and 2 that correspond to less than 5% of contaminated instances classified as non contaminated.

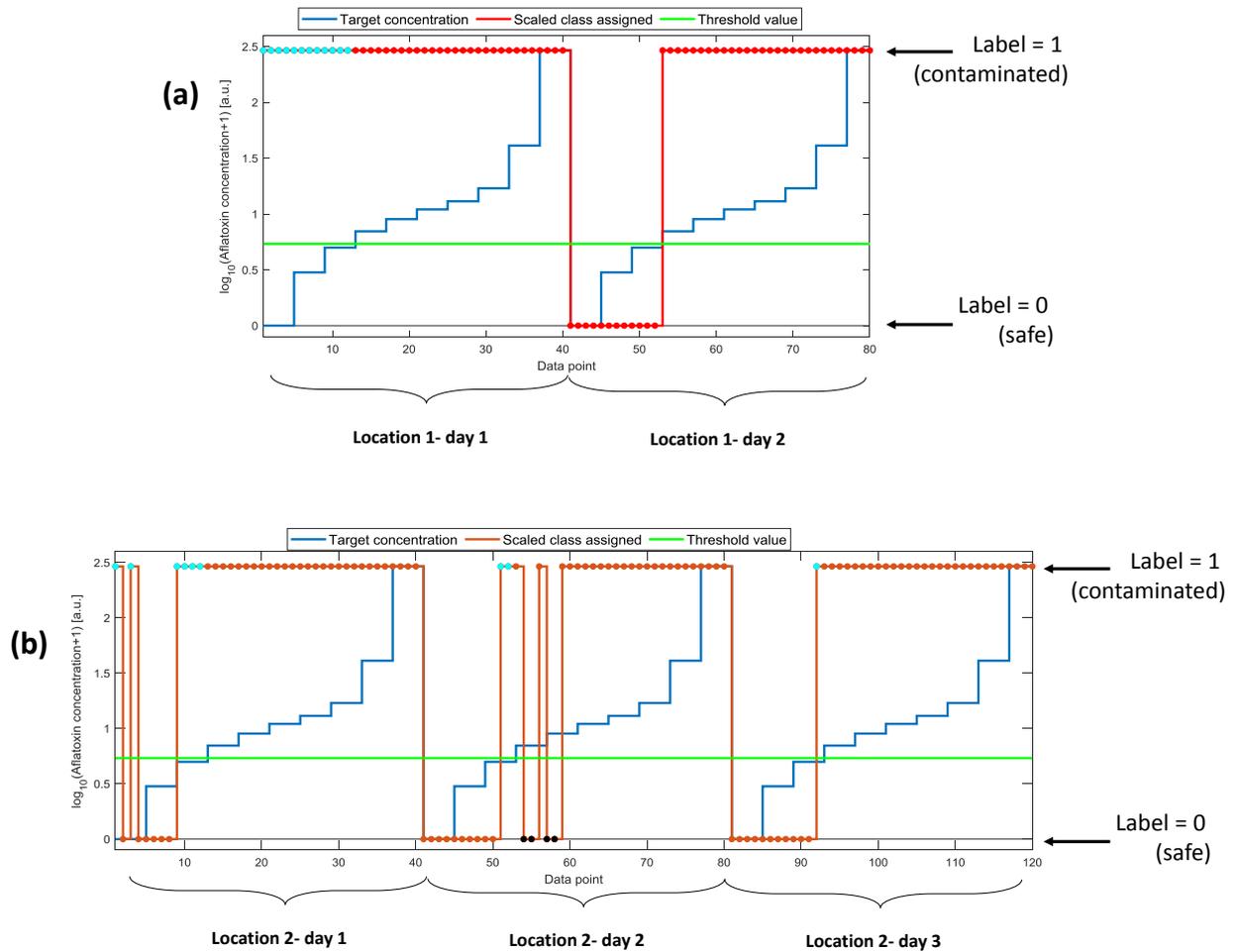

Figure 5. (a) Test at location1 (b) Test at location 2. Blue curves represent the target concentrations. The dashed red curves represent the scaled label assigned by the classifier. The green line is the threshold limit set at 4.4 ng/g. All the values are represented in logarithmic scale for the sake of visibility and grouped after majority voting by sample (and not by point of acquisition). Cyan markers locate the FP samples and black markers identify the FN samples.

We represent in Fig. 5 the labels assigned over the original target concentration for the location 1 (Fig. 5 a) and location 2 (Fig. 5 b) after the application of majority voting grouped by sample. The blue curves represent the actual aflatoxin concentration of the sample. The red dashed curves represent the scaled label assigned to the sample. The green line indicates the threshold limit of 4.4 ng/g. All the values are represented in logarithmic scale ($\log(conc + 1)$) to improve visibility. Cyan and black markers locate the FP and the FN samples respectively. As it can be noted, in experiment a) there are not any FN samples while there are 12 FP samples. In experiment b) there are only 4 FN samples ad 9 FP samples. In particular, FN samples (black markers) exhibit concentrations equal to 6.4 ng/g and 8.9 ng/g that are relatively close to the threshold limit.

# Conclusion

The preliminary data here reported allow us to be confident on the capability of fluorescence approach to detect aflatoxin contamination. In this paper, a binary classification model based on SVM algorithm with Radial Basis Function kernel has been applied after a set of pre-processing steps to select the significant features from the data sets. Moreover, a majority vote procedure has been performed on the classification results coming from spectra acquired on samples with the same reference contamination and acquired in different spatial positions on the same sample. It must be underlined that the analysed data have been collected during two different measurement sessions performed in two different locations. The outcome of the performed analysis is quite encouraging: we could fix a threshold value in the contamination level and assign the measured spectra to the contaminated/non contaminated labels with an accuracy of more than 70% in all the considered conditions.

Anyway, the process of analysis in all its complexity has allowed to identify also the main criticalities of the method used, opening the way to of potential improvements and further developments, the creation of new models taking into account the possible non-uniformity of the natural contamination levels, the employment of the same standardized instrument for all the measures in order to reduce data dispersion.

The proposed spectroscopic technique together with the machine learning analysis shows a great potential and may already represent a helpful approach in developing countries where rapid and low cost contamination testing would be of great benefit (Shephard and Gelderblom, 2014). Importantly, the reported results have been obtained in a very simple set-up in which no focusing of the excitation light has been applied and data analysis performed.

The described results, while quite preliminary, demonstrate the potentiality of such an integrated approach to develop a simple and portable device for food safety assessment. The general interest in smart devices often combined with smartphones is confirmed by recent literature (Hossain et al., 2016; Wilkes et al., 2017) and the development of real time, low cost and easy to use food sensors is of great importance both for end users and for developing countries. Further optical and analytical integration of fluorescence emission and visible reflectance spectra with NIR reflectance spectra, as planned in developing

PhasmaFOOD project, will allow a more detailed description of contaminants presence and could enable shelf life prediction for fresh food (Tsakanikas et al., 2018).

**Acknowledgements**. We acknowledge funding from the EU project PhasmaFOOD (contract n. 732541). Yannick Weesepoel, Judith Mueller-Maatsch and Martin Alewijn at Rikilt laboratories for their support in data acquisition. We thank Dr. F. Epifani for preparing the almond culture material inoculated with *A. flavus* at CNR-ISPA laboratories.